\documentclass[12pt]{article}
\usepackage{graphicx}
\begin{document}

\begin{center}
{\Large \bf The Language of Two-by-two Matrices\\[1mm]
spoken by Optical Devices}

\vspace{7mm}

Y. S. Kim\footnote{electronic address: yskim@umd.edu}\\
Center for Fundamental Physics, University of Maryland,\\
College Park, Maryland 20742

\end{center}

\vspace{6mm}

\begin{abstract}

With its three independent parameters, the $ABCD$ matrix serves as
the beam transfer matrix in optics.  If it is transformed to an
equi-diagonal form, the matrix has only two independent parameters
determined by optical devices.  It is shown that this two-parameter
mathematical device contains enough information to describe the
basic space-time symmetry of elementary particles. If its trace is
smaller than two, this matrix can represent the internal space-time
symmetry of massive particles.  If equal to two, the matrix is of
massless particles. If the trace is greater than two, this matrix
describes imaginary-mass particles.   This matrix speaks Einstein's
language for space-time structure of elementary particles.
As for the optical devices, the laser cavity and the multilayer
system are discussed as illustrative physical examples.

\end{abstract}

\section{Introduction}\label{intro}
The $ABCD$ matrix is a two-by-two matrix with real elements, and
its determinant is one.  There are therefore three independent
parameters.  These elements are determined by optical materials
and how they are arranged.  The purpose of this note is to explore
its mathematical properties which can address more fundamental
issues in physics.
\par
First of all, the trace of this matrix could be less than two, equal
to two, or greater than two.  We are interested in what physical
conclusions we can derive from these numbers.
\par
In order to bring the $ABCD$ matrix to the form which will address
some fundamental issues in physics, we should first transform it into
the equi-diagonal form where the two-diagonal elements are equal to
each other~\cite{bk09josa,bk10jmo}.  We can achieve this goal by a
similarity transformation with a one-parameter matrix.  This
transformation does not change the trace, and the resulting
equi-diagonal matrix has two independent parameters.
\par
We shall call this equi-diagonal matrix the core of the $ABCD$
matrix, and use the notation $[ABCD]$.  This matrix cannot always be
diagonalized.  This creates a non-trivial problem.  We shall examine
how optical devices, especially periodic systems, can lead us to a
better understanding of the problem.  For this purpose, we discuss
laser cavities and multilayer systems in detail.
\par
If the trace is less than two, the core can be written as
\begin{equation}\label{core01}
[ABCD] =\pmatrix{\cos(\gamma/2) & - e^{\eta} \sin(\gamma/2) \cr
  e^{-\eta} \sin(\gamma/2)  & \cos(\gamma/2)} .
\end{equation}
The diagonal elements are equal and smaller than one.
\par
If the trace is greater than two, the $[ABCD]$ matrix takes the
form
\begin{equation}\label{core02}
[ABCD] = \pmatrix{\cosh(\gamma/2) & e^{\eta} \sinh(\gamma/2) \cr
  e^{-\eta} \sinh(\gamma/2)  & \cosh(\gamma/2)} .
\end{equation}
Here again the diagonal elements are equal, but they are greater
than one.
\par
If the trace is equal to two, the core matrix becomes
\begin{equation}\label{core03}
[ABCD] = \pmatrix{1 &  -\gamma \cr 0  & 1 } .
\end{equation}
This matrix also has the same diagonal element, and they are equal
to one.

\par
This triangular matrix of Eq.(\ref{core03}) cannot be diagonalized.
The core matrices of Eq.(\ref{core01}) and Eq.(\ref{core02}) can
be diagonalized, but not by rotation alone.  These mathematical
subtleties are not well known.  The purpose of this report is to
show how much physics we can extract from these mathematical details.

\par
The mathematics of group theory allows us to write down a four-by-four
Lorentz-transformation matrix for every two-by-two matrix discussed
in this paper.  In this way, the three matrices given in
Eq.(\ref{core01}), and Eq.(\ref{core02}), and Eq.(\ref{core03}) lead
to the internal space-time symmetries of elementary particles.
They respectively correspond to the symmetries of massive,
imaginary-mass, and massless particles respectively~\cite{wig39}.

\par
In Sec.~\ref{decomp}, we write the $[ABCD]$ matrix in terms of the
its generators, and decompose it to three matrices in the form of
similarity transformation.  It is noted that there is another
mathematical device known as the Bargmann decomposition~\cite{barg47}.
This decomposition is not a similarity transformation, but it
can play other important roles in understanding the $ABCD$ matrix.
\par
In Sec.~\ref{lacav}, we discuss a laser cavity as a physical
example of the mathematical details of the $[ABCD]$ matrix.  In
Sec.~\ref{multi}, a multilayer system is discussed as a physical
example leading to the desired similarity transformation.  The
Bargmann decomposition plays the key role in this problem.
In Sec.~\ref{sptime}, we shall discuss the space-time symmetries
implied by the properties of the $[ABCD]$ matrix.  The Bargmann
decomposition, as well the similarity decompositions, is explained
in terms of the Lorentz transformations which leave the momentum
of a given particle invariant.  Thus, these transformations are
applicable to internal space-time symmetries.

\section{Decomposition of the ABCD Matrix}\label{decomp}
We are interested in writing the three different forms of
the core matrix in one expression.
\begin{equation}\label{core10}
[ABCD] = \exp{\left\{\frac{1}{2}
   \pmatrix{0 & - x - y  \cr x - y & 0}\right\}} ,
\end{equation}
where the parameters $x$ and $y$ are determined by the optical
materials and how they are arranged.  The exponent of this matrix
becomes
\begin{equation}
\frac{1}{2} \pmatrix{0 & - x - y  \cr x - y & 0} .
\end{equation}

\par

If $x > y $, the matrix becomes
\begin{equation}
\frac{\gamma}{2} \pmatrix{0 & -\exp{(\eta/2)}  \cr \exp{(-\eta/2)} & 0} ,
\end{equation}
which leads to the core matrix of Eq.(\ref{core01}) with
\begin{eqnarray}\label{core11}
&{}& \gamma = \sqrt{x^2 - y^2}, \nonumber\\[1ex]
&{}& e^{\eta} = \sqrt{\frac{x + y}{x - y}} .
\end{eqnarray}
The core matrix $[ABCD]$ can be written as a similarity transformation
\begin{equation}\label{core16}
[ABCD] = B(\eta) R(\theta) B(-\eta)
\end{equation}
with
\begin{eqnarray}\label{core26}
&{}& B(\eta) = \pmatrix{e^{\eta/2} & 0 \cr 0 & e^{-\eta/2}} \nonumber \\[1ex]
&{}& R(\theta) = \pmatrix{\cos(\theta/2) & -\sin(\theta/2)
     \cr \sin(\theta/2) & \cos(\theta/2)} ,
\end{eqnarray}
where $\gamma$ is now replaced by the rotation angle $\theta$.
$R(\theta)$ is a rotation matrix, and $B(\eta)$ is a squeeze matrix.

\par
If $x < y $, the exponent becomes
\begin{equation}
\frac{\gamma}{2}
   \pmatrix{0 & \exp{(\eta/2)}  \cr \exp{(-\eta/2)} & 0},
\end{equation}
leading to the core matrix of Eq.(\ref{core02}), with
\begin{eqnarray}\label{core12}
&{}& \gamma = \sqrt{y^2 - x^2}, \nonumber\\[1ex]
&{}& e^{\eta} = \sqrt{\frac{x + y}{y - x}} .
\end{eqnarray}
The $[ABCD]$ matrix can now be decomposed into a similarity transformation
\begin{equation}\label{bargw11}
[ABCD] = B(\eta) S(-\lambda) B(-\eta) ,
\end{equation}
with
\begin{equation}\label{core36}
S(\lambda) = \pmatrix{\cosh(\lambda/2) & \sinh(\lambda/2)
     \cr \sin(\lambda/2) & \cosh(\lambda/2)} ,
\end{equation}
where $\gamma$ is replaced by the boost parameter $\lambda$.  The
matrix $B(\eta)$ takes the diagonal form given in Eq.(\ref{core16})
with $\eta$ defined in Eq.(\ref{core12}).  $S(\lambda)$ is a squeeze
matrix.
\par
If $x = y$, the exponent becomes
\begin{equation}\label{core55}
  \pmatrix{1 &  - x \cr 0  & 1 } ,
\end{equation}
with $x = y = \gamma$.
\par
We now have combined three different expressions for the core of the
$ABCD$ matrix into one exponential form of Eq.(\ref{core10}). This
form can be decomposed into three matrices constituting a similarity
transformation.  We shall call this the ``Wigner decomposition'' for
the reasons given in Sec.~\ref{sptime}.

\par
There is another form of decomposition known as the Bargmann
decomposition~\cite{barg47}, which states that the core of the $ABCD$
matrix can be written as
\begin{equation}\label{barg11}
[ABCD] = R(\alpha) S(-2\chi) R(\alpha) ,
\end{equation}
where the forms of the rotation matrix $R$ and the squeeze matrix $S$
are given in Eq.(\ref{core16}) and Eq.(\ref{core26}) respectively.
If we carry out the matrix multiplication, the $[ABCD]$ matrix becomes
\begin{equation}\label{barg22}
\pmatrix{(\cosh\chi)\cos\alpha  &
       -\sinh\chi - (\cosh\chi)\sin\alpha  \cr
      -\sinh\chi + (\cosh\chi)\sin\alpha  &
  (\cosh\chi)\cos\alpha } .
\end{equation}
This matrix also has two independent parameters $\alpha$ and $\chi$.
We can write these parameters in terms of $\gamma$ and $\eta$ by
comparing the matrix elements.  For instance, if $x > y$, the
diagonal elements lead to
\begin{equation}
\cos(\theta/2) = (\cosh\chi) \cos\alpha .
\end{equation}
The off-diagonal elements lead to
\begin{equation}
e^{2\eta} = \frac{(cosh\chi)\sin\alpha + \sinh\chi}
           {(cosh\chi)\sin\alpha - \sinh\chi}
\end{equation}

\par

As for physical applications, let us consider periodic systems,
such as laser cavities and multilayer systems.  The exponential
form given in Eq.(\ref{core10}) tells us that it is a matter of
replacing the $\gamma$ parameter by $N\gamma$ for N repeated
applications~\cite{bk10jmo}.  Let us see some examples.
\par

\section{Laser Cavities}\label{lacav}
As the first example, let us consider the laser cavity consisting
of two identical concave mirrors separated by a distance $d$.
Then the $ABCD$ matrix for a round trip of one beam is
\begin{equation}\label{cav11}
  \pmatrix{1 & 0 \cr -2/R & 1}
  \pmatrix{1 & d \cr 0 & 1}
  \pmatrix{1 & 0 \cr -2/R & 1} \pmatrix{1 & d \cr 0 & 1},
\end{equation}
where
the matrices
\begin{equation}\label{rad}
  \pmatrix{1 & 0 \cr -2/R & 1} ,  \quad
  \pmatrix{1 & d \cr 0 & 1}
\end{equation}
are the mirror and translation matrices respectively.  The parameters
$R$ and $d$ are the radius of the mirror and the mirror separation
respectively.  This form is quite familiar to us from the laser
literature~\cite{yariv75,haus84,hawk95,saleh07}.
\par
However, the crucial question is what happens when this process is
repeated.  We are thus led to the question of whether the chain of
matrices in Eq.(\ref{cav11}) can be brought to an equi-diagonal form
and eventually to a form of the Wigner decomposition.  We are
interested in finding the core of Eq.(\ref{cav11}).
For his purpose,  we rewrite the matrix of Eq.(\ref{cav11}) as
\begin{eqnarray}\label{abcd3}
&{}&  \pmatrix{1 & -d/2 \cr 0 & 1}
  \pmatrix{1 & d/2 \cr 0 & 1}
  \pmatrix{1 & 0 \cr -2/R & 1}
  \pmatrix{1 & d/2 \cr 0 & 1}^{2} \nonumber\\[1ex]
&{}&  \times
  \pmatrix{1 & 0 \cr -2/R & 1}
  \pmatrix{1 & d/2 \cr 0 & 1}
  \pmatrix{1 & d/2 \cr 0 & 1} .
\end{eqnarray}

\par
In this way, we translate the system by $-d/2$ using a translation
matrix given in Eq.(\ref{rad}), and  write the
$ABCD$ matrix of Eq.(\ref{cav11}) as
\begin{equation}
\pmatrix{1 & -d/2 \cr 0 & 1}
\left[\pmatrix{1 - d/R &   d - d^{2}/2R  \cr -2/R & 1 - d/R}\right]^{2}
\pmatrix{1 & d/2 \cr 0 & 1}.
\end{equation}
We are thus led to concentrate on the matrix in the middle
\begin{equation}
 \pmatrix{1 - d/R &   d - d^{2}/2R  \cr -2/R & 1 - d/R},
\end{equation}
which can be written as
\begin{equation}
 \pmatrix{\sqrt{d} & 0  \cr 0 & 1/\sqrt{d}}
  \pmatrix{1 - d/R &   1 - d/2R  \cr -2d/R & 1 - d/R}
  \pmatrix{1/\sqrt{d} & 0  \cr 0 & \sqrt{d}} .
\end{equation}
It is then possible to decompose the $ABCD$ matrix into
\begin{equation}
E~C^{2}~E^{-1} ,
\end{equation}
with
\begin{eqnarray}\label{escort}
&{}&  C = \pmatrix{1 - d/R &  1 - d/2R  \cr
         -2d/R & 1 - d/R},  \nonumber \\[2mm]
&{}&  E = \pmatrix{1 & -d/2 \cr 0 & 1}
\pmatrix{\sqrt{d} & 0 \cr 0 & 1/\sqrt{d}} .
\end{eqnarray}
The $C$ matrix now contains only dimensionless numbers, and it
can be written as
\begin{equation}
C = \pmatrix{\cos(\gamma/2) & e^{\eta}\sin(\gamma/2)  \cr
    - e^{-\eta} \sin(\gamma/2)  & \cos(\gamma/2)} ,
\end{equation}
with
\begin{eqnarray}
&{}&\cos(\gamma/2) = 1 - \frac{d}{R},   \nonumber\\[1ex]
&{}& e^\eta = \sqrt{\frac{2R - d}{4d}}
\end{eqnarray}
Here both $d$ and $R$ are positive, and the restriction on them is
that $d$ be smaller than $2R$.  This is the stability condition
frequently mentioned in the literature~\cite{haus84,hawk95}.
\par

Thus, the $[ABCD]$ core matrix is $C^2$, and takes the form
\begin{equation}
[ABCD]  = \pmatrix{\cos(\gamma) & e^{\eta}\sin(\gamma)  \cr
    - e^{-\eta} \sin(\gamma)  & \cos(\gamma)} ,
\end{equation}
and the similarity transformation which connects this core matrix
with the original $ABCD$ matrix of Eq.(\ref{cav11}) is the matrix
$E$ given in Eq.(\ref{escort}).

\section{Multilayer Optics}\label{multi}
We consider an optical beam going through a periodic medium with
two different refractive indexes.   If the beam traveling in the
first medium hits the second medium, it is partially transmitted
and partially reflected.  In order to maintain the continuity
in the pointing picture, we normalize the electric field as
\begin{eqnarray}
E_{1}^{(\pm)} = \frac{1}{\sqrt{n_1}}
       \exp{\left({\pm}ik_1 z  - \omega t \right)} \nonumber\\[1ex]
E_{2}^{(+)} = \frac{1}{\sqrt{n_2}}
           \exp{\left({\pm}ik_2 z  - \omega t \right)}
\end{eqnarray}
for the optical beams in the first and second media respectively.
The superscript $(+)$ and $(-)$ are for the incoming and reflected
rays respectively.
\par
These two optical rays are related by the two-by-two $ABCD$ matrix,
according to
\begin{equation}\label{smat}
\pmatrix{E_{2}^{(+)} \cr E_{2}^{(-)}} =
\pmatrix{A & B \cr C & D}
\pmatrix{E_{1}^{(+)} \cr E_{1}^{(-)}} .
\end{equation}
Of course the elements of this matrix are determined by
transmission coefficients as well as the phase shifts the beams
experience while going through the media~\cite{azzam77}.
\par
When the beam goes through the first medium to the second, we
may use the boundary matrix\cite{monzon00}
\begin{equation}\label{boundary11}
Q(\sigma) = \pmatrix{\cosh(\sigma/2) & \sinh(\sigma/2) \cr
              \sinh(\sigma/2) & \cosh(\sigma/2) } ,
\end{equation}
where the parameter $\sigma$ is determined by the reflection and the
transmission coefficients~\cite{saleh07,azzam77,monzon00}.  Then the
boundary matrix for the beam going from the second medium
should be $Q(-\sigma)$.
\par
In addition, we have to consider the phase shifts the beams have to
go through.  When the beam goes trough the first media, we can use
the phase-shift matrix
\begin{equation}
P\left(\delta_1\right) =
\pmatrix{e^{-i\delta_1/2} &  0 \cr 0 & e^{i\delta_1/2} }   ,
\end{equation}
and a similar expression for $P\left(\delta_2\right)$ for the second
medium.
\par
We are thus led to consider one complete cycle starting from the midpoint
of the second medium, and write
\begin{equation} \label{chain11}
 P\left(\delta_2/2\right) Q(\sigma) P\left(\delta_1\right) Q(-\sigma)
  P\left(\delta_2/2\right) .
\end{equation}
\par

If multiplied into one matrix, is this matrix equi-diagonal to accept
the Wigner and Bargmann decompositions?  Another question is whether
the matrices in the above expression can be converted into matrices
with real element.

In order to answer the second question, let us consider the similarity
transformation
\begin{equation}\label{conju66}
C_1~P(\delta) Q(\sigma)~C_1^{-1} ,
\end{equation}
with
\begin{equation}
C_{1} =  {1 \over \sqrt{2}} \pmatrix{1 & i \cr i & 1} .
\end{equation}
This transformation leads to
\begin{equation} \label{conju77}
  R(\delta) Q(\sigma) ,
\end{equation}
where
\begin{equation}
R(\delta) = \pmatrix{\cos(\delta/2) & -\sin(\delta/2) \cr
                     \sin(\delta/2) & \cos(\delta/2) } .
\end{equation}
This notation is consistent with the rotation matrices
used in Sec.~\ref{decomp}.

\par
Let us make another similarity transformation with
\begin{equation}
C_{2} =  {1 \over \sqrt{2}} \pmatrix{1 & 1 \cr -1 & 1} .
\end{equation}
This changes $Q(\sigma)$ into $B(\sigma)$ without changing
$R(\delta)$,
where
\begin{equation}
B(\sigma) = \pmatrix{e^{\sigma/2} & 0 \cr 0 & e^{-\sigma/2}},
\end{equation}
again consistent with the $B(\eta)$ matrix used in
Sec.~\ref{decomp}.

 \par

Thus the net similarity transformation matrix is~\cite{gk01}
\begin{equation}\label{ccc}
C = C_{2}C_{1} = {1 \over \sqrt{2}} \pmatrix{e^{i\pi/4} &  e^{i\pi/4}
\cr -e^{-i\pi/4} & e^{-i\pi/4}} ,
\end{equation}
with
\begin{equation}
C^{-1} = {1 \over \sqrt{2}} \pmatrix{e^{-i\pi/4} &  -e^{i\pi/4} \cr
e^{-i\pi/4} & e^{i\pi/4}} .
\end{equation}

\par

If we apply this similarity transformation to the long matrix chain of
Eq.(\ref{chain11}), it becomes another chain
\begin{equation} \label{chain22}
M = R\left(\delta_2/2\right) B(\sigma) R\left(\delta_1\right) B(-\sigma)
                          R\left(\delta_2/2\right) ,
\end{equation}
where all the matrices are real.
\par

Let us now address the main question of whether this matrix chain can
be brought to one equi-diagonal matrix.  We note first that the
three middle matrices can be written in a familiar form
\begin{eqnarray} \label{chain33}
&{}& \hspace{-5mm} M =  B(\sigma) R\left(\delta_1\right) B(-\sigma) \nonumber \\[1ex]
&{}& = \pmatrix{\cos(\delta_1/2) & -e^{\sigma}\sin(\delta_1/2) \cr
 e^{-\sigma}\sin(\delta_1/2)& \cos(\delta_1/2)} .
\end{eqnarray}
However, due to the rotation matrix $R\left(\delta_2/2\right)$ at the
beginning and at the end of Eq.(\ref{chain22}), it is not clear whether the
entire chain can be written as a similarity transformation.
\par
In order to resolve this issue, let us write Eq.(\ref{chain33}) as a
Bargmann
decomposition
\begin{equation}\label{barg66}
R(\alpha) S(-2\chi) R(\alpha) ,
\end{equation}
with its explicit expression given in Eq.(\ref{barg22}).  The
parameters $\alpha$ and $\chi$ are related to $\sigma$ and
$\delta_1$ by
\begin{eqnarray}
&{}& \cos(\delta_1/2) = (\cosh\chi) \cos\alpha ,  \nonumber \\[1ex]
&{}& e^{2\sigma} = \frac{(cosh\chi)\sin\alpha + \sinh\chi}
           {(cosh\chi)\sin\alpha - \sinh\chi} .
\end{eqnarray}
\par
It is now clear that the entire chain of Eq.(\ref{chain11}) can
be written as another Bargmann decomposition
\begin{equation}\label{barg77}
M = R(\alpha + \delta_2/2) S(-2\chi) R(\alpha + \delta_2/2) .
\end{equation}
Finally, if the trace of this matrix is less than two, it can be
converted to a Wigner decomposition
\begin{equation}\label{barg88}
 M = B(\eta) R(\theta) B(-\eta)
\end{equation}
with
\begin{eqnarray}
&{}& \cos(\theta/2) = (\cosh\chi) \cos(\alpha + \delta_2/2),  \nonumber \\[1ex]
&{}& e^{2\eta} = \frac{(\cosh\chi)\sin(\alpha + \delta_2/2) + \sinh\chi}
           {(\cosh\chi)\sin(\alpha + \delta_2/2) - \sinh\chi} .
\end{eqnarray}
Similar expressions can be derived for other values of the trace~\cite{gk03}.
\par
It is interesting to note that the Bargmann decomposition plays an essential
role in bringing the chain of five matrices given in Eq.(\ref{chain22}) to
the Wigner decomposition of Eq.(\ref{barg88}) consisting three matrices.
We shall see what other function the Bargmann decomposition can perform
in Sec.~\ref{sptime}.

\section{Space-time Symmetries}\label{sptime}
These properties are applicable to many other branches of physics.
For instance, one of the  persisting problems is the internal space-time
symmetry of elementary particles in Einstein's Lorentz-covariant world.
\par
The mathematics of group theory allows us to translate the rotation
and squeeze matrices of Eq.(\ref{core16}) and Eq.(\ref{core26})
into the following four-by-four matrices respectively.
\begin{eqnarray}\label{4by4}
&{}& R(\theta) = \pmatrix{\cos\theta & 0 & \sin\theta & 0 \cr
  0 & 1 & 0 & 0 \cr
  -\sin\theta & 0 & \cos\theta &  0 \cr
  0 & 0 & 0 & 1} ,                 \nonumber\\[1ex]
&{}& S(\lambda) = \pmatrix{\cosh\lambda & 0 & 0 & \sinh\lambda \cr
  0 & 1 & 0 & 0 \cr
  0 & 0 & 1 &  0 \cr
  \sinh\lambda & 0 & 0 & \cosh\lambda} , \nonumber\\[1ex]
&{}& B(\eta) = \pmatrix{1 & 0 & 0 & 0\cr
  0 & 1 & 0 & 0 \cr
  0 & 0 & \cosh\eta &  \sinh\eta \cr
  0 & 0 & \sinh\eta & \cosh\eta} ,
\end{eqnarray}

\par
They are applicable to the Minkowskian four-vector $(x, y, z, t)$.
The $R(\theta)$ matrix performs a rotation around the $y$ axis, and
$S(\lambda)$ is for Lorentz boosts along the $x$ axis.
The $B(\eta)$ matrix boosts the system along the $z$ direction.

\par
Together with a rotation matrix around $z$ axis
\begin{equation}\
Z(\phi) = \pmatrix{\cos\phi & -\sin\phi & 0 & 0\cr
  \sin\phi & \cos\phi & 0 & 0 \cr
  0 & 0 & 1 &  0 \cr  0 & 0 & 0 & 1} ,
\end{equation}
they constitute Wigner's little groups dictating internal space-time
symmetries of massive and imaginary-mass particles~\cite{wig39}.
The triangular matrix of Eq.(\ref{core03}) leads to the little group
for massless particles.  The little groups are the subgroups of the
Lorentz group whose transformations leave the four-momentum of a given
particle invariant.
\par

Let us go back to Eq.(\ref{core01}) which,  according to Eq.(\ref{core16}),
can be decomposed to a similarity transformation
\begin{equation}\label{sim44}
   W(\eta, \theta) = B(\eta) R(\theta) B(-\eta).
\end{equation}
We can write this decomposition with the four-by four matrices given in
in Eq.(\ref{4by4}).
\par
Let us then consider a massive particle moving along the $z$
direction with the velocity parameter $v/c = \tanh\eta$, and its
four-momentum
\begin{equation}
 (0, 0, m\sinh\eta, m\cosh\eta),
\end{equation}
where $m$ is the mass of the particle.
\par
\begin{figure}
\centerline{\includegraphics[scale=0.43]{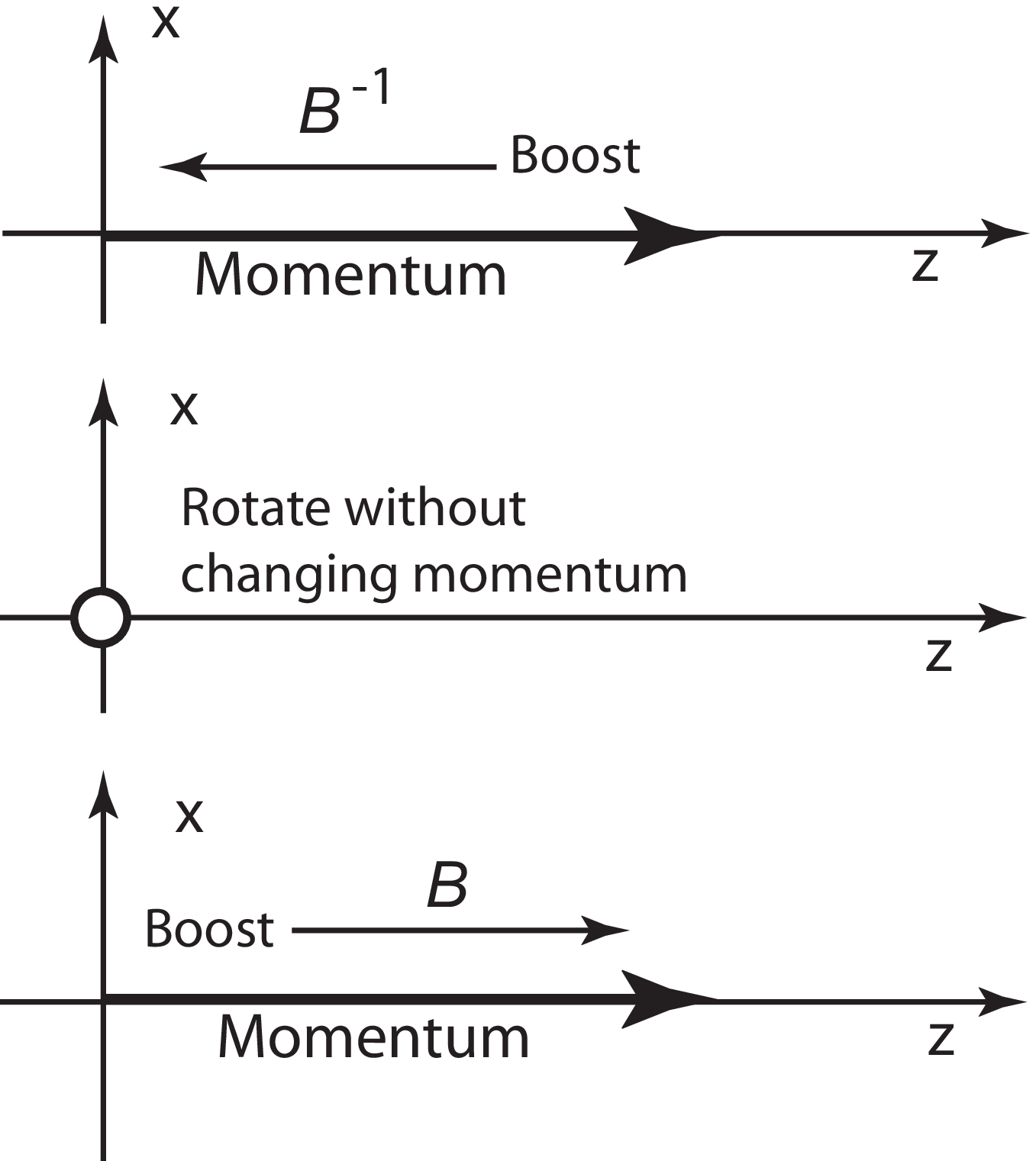}
\hspace{10mm} \includegraphics[scale=0.34]{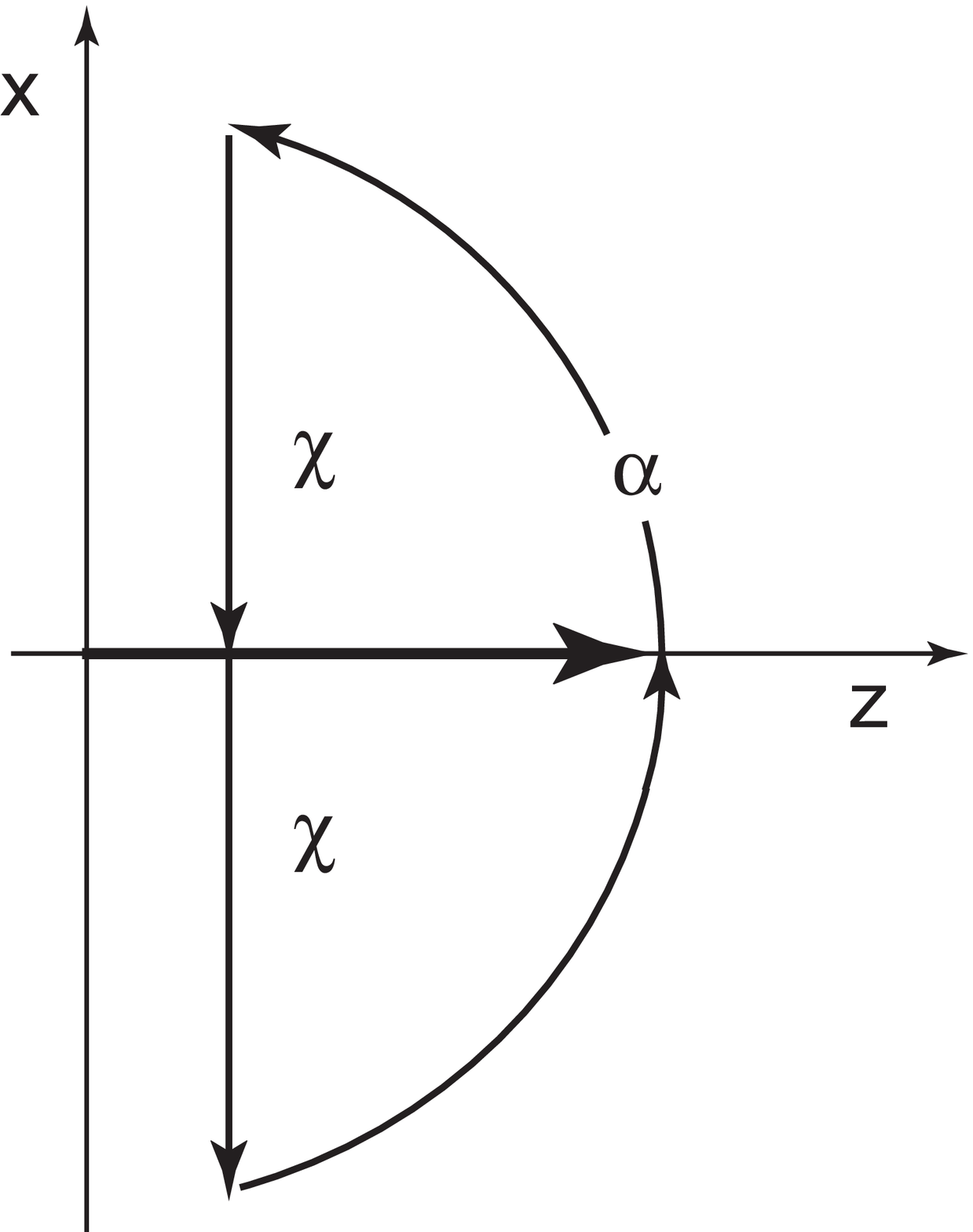}}
\caption{Wigner decomposition (left) and Bargmann decomposition (right).
These figures illustrate momentum preserving transformations.  In the
Wigner transformation, a massive particle is brought to its rest frame.
It can be rotated while the momentum remains the same.  This particle
is then boosted back to the frame with its original momentum.  In the
Bargmann decomposition, the momentum is rotated, boosted, and rotated
to its original position.}\label{bargwig}
\end{figure}

We can boost this particle using the boost matrix $B(-\eta)$, which
is the inverse of the four-by-four matrix  given in Eq.(\ref{4by4}).
The particle becomes at rest, with its four-momentum
\begin{equation}
 (0, 0, 0, m),
\end{equation}
and with zero velocity.  The rotation matrix $R(\theta)$ rotates
this particle without changing its momentum.  During this process,
the particle changes the direction of its spin.  Finally, $B(\eta)$
boosts the particle and restores its momentum, as is illustrated
in Fig.~\ref{bargwig}.  Thus, it is appropriate to call
the form of Eq.(\ref{sim44}) the Wigner decomposition.
In this way, the four-by-four expression for Eq.(\ref{core16})
changes the internal space-time structure of the particle.
\par
The essential function of the Wigner decomposition is to provide
the subgroup of the Lorentz group which will leave the given
four-momentum of a particle invariant~\cite{wig39}.
\par
The Bargmann decomposition also provides momentum-preserving
transformations. The decomposition consists of a rotation,
a boost, and another rotation as illustrated in Fig.~\ref{bargwig}.
It is interesting to note that the Wigner decomposition and
the Bargmann decomposition can serve the same purpose of providing
a Lorentz transformation which leaves the momentum invariant.
\par

One key question from this table is what happens to the $O(3)$-like
little group when the particle momentum becomes infinity or its mass
becomes zero. The question is whether the little group for a massive
particles becomes that for a massless particle.  The answer to this
question is Yes, but this issue had a stormy history before this
definitive answer~\cite{kiwi90jmp}.  Indeed, when $\eta$ becomes
infinity, the four-by-four form of Eq(\ref{sim44}) becomes
\begin{equation}
\pmatrix{1 & 0 & -\gamma &\gamma \cr  0 & 1 & 0 & 0   \cr
\gamma & 0 & 1 - \gamma^2/2  &  \gamma^2/2 \cr
\gamma & 0  & -\gamma^2/2    & 1 + \gamma^2/2 } .
\end{equation}
When applied to the momentum of a massless particle
moving in the negative $z$ direction with
\begin{equation}
(0, 0, p, p),
\end{equation}
This matrix leaves the above four-momentum invariant, but it
performs a gauge transformation~\cite{kiwi90jmp}.  This aspect
of Wigner's little group is illustrated in Table~\ref{eiwi}.

\par

\begin{table}[thb]
\caption{Massive and massless particles in one package.  Einstein
unified the energy-momentum relation for slow (massive) and fast
(massless) particles with one Lorentz-covariant formula.  Likewise,
Wigner's little group unifies the internal space-time symmetries of
particles.  When boosted, the spin along the direction remains
invariant.   This is called the Lorentz-invariant helicity.  The
spins along the transverse directions collapse into a gauge degree
of freedom in the infinite-momentum or zero-mass
limit~\cite{kiwi90jmp,hks86jmp}.}\label{eiwi}

\begin{center}
\begin{tabular}{lccc}
\hline
\hline\\[1mm]
{} & Massive & Lorentz & Massless \\
{} & Slow  & Covariance & Fast \\[2mm]
\hline
{}&{}&{}&{}\\
 Energy- & $E =$   & Einstein's & {} \\
Momentum &  $p^{2}/2m$ & $ E = [p^{2} + m^{2}]^{1/2}$ & $E = p$ \\[4mm]
\hline
{}&{}&{}&{}\\
Internal & $S_3$  & Wigner's  & $S_3$ \\
Symmetries & $S_1, S_2$  & Little Group & Gauge Trans.  \\[4mm]
\hline
\hline
\end{tabular}
\end{center}
\end{table}

\newpage
\section*{Concluding Remarks}

In this report, we have discussed some properties of the $ABCD$
matrix which serves as the standard research tool in ray optics.
This two-by-two matrix four real elements, but only three
independent parameters if the determinant of the matrix is
constrained to be one.
\par
If the determinant is restricted to be one, the $ABCD$ matrix has
three independent parameters.  It was noted that this matrix can be
written as a similarity transformation of a core matrix with two
independent parameters.  Two physical examples are given to
illustrate how these parameters are determined from optical devices.
\par
It is remarkable that this two-parameter matrix contains enough
information to describe the internal space-time structure of
elementary particles.

\end{document}